\documentstyle[12pt]{article}
\def \bea{\begin{eqnarray}}
\def \beq{\begin{equation}}
\def \eea{\end{eqnarray}}
\def \eeq{\end{equation}}
\def \ite{{\it et al.}}
\def \s{\sqrt{2}}

\def \siml{\stackrel{<}{\sim}}
\def \st{\sqrt{3}}
\def \sx{\sqrt{6}}
\begin{document}
\renewcommand{\thetable}{\Roman{table}}
\rightline{TECHNION-PH-99-33}
\rightline{EFI 99-40}
\rightline{September 1999}
\rightline{hep-ph/9909478}
\bigskip

\centerline{\bf NEW INFORMATION ON $B$ DECAYS}
\centerline{\bf TO CHARMLESS $VP$ FINAL STATES
\footnote{Submitted to Phys.~Rev.~D.}}
\bigskip
\centerline{\it Michael Gronau}
\centerline{\it Physics Department}
\centerline{\it Technion -- Israel Institute of Technology, 32000 Haifa,
Israel}
\smallskip
\centerline{and}
\smallskip
\centerline{\it Jonathan L. Rosner}
\centerline{\it Enrico Fermi Institute and Department of Physics,
University of Chicago}
\centerline{\it 5640 S. Ellis Avenue, Chicago IL 60615}
\bigskip

\centerline{\bf ABSTRACT}
\bigskip

\begin{quote}
The decays of $B$ mesons to charmless final states consisting of a vector meson
($V$) and a pseudoscalar meson ($P$) are analyzed within flavor SU(3).
Predictions are compared with new data from the CLEO Collaboration. Dominant
contributions to amplitudes and subdominant interfering terms are identified.
Evidence is found for a specific penguin amplitude (contributing, for example,
to $B^+ \to \rho^+ K^0$) at a level much higher than that implied by most
explicit models. The validity of the conclusion $\cos \gamma < 0$, obtained
through other analyses of $B \to VP$ decays, is examined here from a less
model-dependent standpoint. It is found that several processes are consistent
with $\cos \gamma < 0$ (or $\cos \alpha > 0$), and measurements are suggested
which could make this conclusion more robust. 
\end{quote}
\bigskip
\leftline{PACS numbers: 13.25.Hw, 14.40.Nd, 11.30.Er, 12.15.Ji}
\bigskip

\centerline{\bf I.  INTRODUCTION}
\bigskip

The decays of $B$ mesons to charmless final states consisting of a vector meson
($V$) and a pseudoscalar meson ($P$) are of potentially great interest in the
study of the weak interactions and CP violation.  The decays $B^0 \to \rho^\mp
\pi^\pm$ occur with a substantially greater combined branching ratio than $B^0
\to \pi^+ \pi^-$, and hence may be useful in CP studies \cite{RP}. The decays
$B^+ \to \omega \pi^+$, $B \to \omega K $, and $B \to \rho K$ can shed light on
decay mechanisms by validating or falsifying specific models (most of which
predict very low rates for $B \to (\omega,\rho) K$).  Decays involving $\eta$
and $\eta'$ mesons are interesting because there exist models for the large
observed rate for $B \to K \eta'$ which make specific predictions for $B \to
K^* \eta$ and $B \to K^* \eta'$. 

In Ref.~\cite{DGR} we made a first attempt to classify $B \to VP$ decays in a
model-independent manner, using only flavor SU(3) symmetry as expressed by a
set of reduced amplitudes depicted in graphical form.  The present article is
an update of that work in the light of new experimental results from the CLEO
Collaboration working at the Cornell Electron Storage Ring (CESR).  These
results include new branching ratios for charmless hadronic $B$ decays to $VP$,
where $V = K^*$, $\rho$, $\omega$, and $\phi$, and $P = (\pi,~K)$ \cite{CLEOVP}
or $B \to K^* (\eta,\eta')$ \cite{CLEOeta}.  We shall also make use of some new
results on $B \to PP$ decays \cite{CLEOPP}, and will use the fact that no
charge asymmetries have been seen in several processes \cite{CLEOasy}. 

One of the issues to which the new data are relevant is the phase of the
Cabibbo-Kobayashi-Maskawa (CKM) matrix element $V_{ub}$:  $\gamma = {\rm
Arg}(V_{ub}^*)$ in a standard convention.  Data on $B \to K \pi$ decays have
been used for several years to constrain this phase, both in relatively
model-independent analyses \cite{GRL,FM,GRgamma,NRPL,NRPRL,MN,GPY,GP} and with
the help of the factorization ansatz and models for form factors
\cite{CLEOPP,HHY,AKL,HSW,CY}. These studies tended to favor $\cos \gamma <0$,
consistent with determinations of CKM parameters which take realistic account
of theoretical errors \cite{Falk} but in some conflict with more optimistic
estimates of these errors \cite{Parodi}.  New factorization-based studies
\cite{HHY,HY} also find evidence for $\cos \gamma <0$ in a number of $B \to VP$
processes.  We wish to determine whether this conclusion holds in a less
model-dependent context. 

We find that {\it if final-state phase shifts are small}, consistent with the
upper limits on charge asymmetries in several $B$ decays to charmless final
states \cite{CLEOasy}), then $B \to VP$ decays indeed favor $\cos \gamma < 0$
(or $\cos \alpha >0$).  Using the model-independent amplitudes obtained from
the $B \to PV$ processes studied by CLEO, we can identify decay modes whose
discovery or improved measurement would permit conclusions about $\cos \gamma$
and other CKM phases to be placed on a firmer footing.  To exhibit the minimal
experimental requirements for demonstrating interference-based constraints on
$\cos \gamma$ or $\cos \alpha$, we perform this analysis independently of other
constraints on CKM phases, mentioning them at the end of this paper. 

A further result we obtain, in contrast to most explicit models, is that the
penguin contributions to the decays $B \to (\omega,\rho) K$ are appreciable. In
particular, the decay $B^+ \to \rho^+ K^0$ should be observable with a
branching ratio in excess of $5 \times 10^{-6}$. 

In Section II we recall some notation from Ref.~\cite{DGR} and tabulate
experimental branching ratios and limits.  Experimental results are taken from
Ref.~\cite{CLEOVP} unless not given there, in which case earlier limits
\cite{oldVP} are quoted.  

The crude nature of present experimental data on $B \to VP$ decays requires
that a phenomenological analysis such as ours proceed in a somewhat roundabout
way.  Instead of simply presenting the results of an overall fit, which would
fail to highlight the places where improved data are necessary, we focus first
(in Sec.~III) on the main amplitudes for which experimental data exist.

In Sec.~IV we then see if deviations from rate relations predicted
in the presence of a single amplitude can detect interference with one or more
subdominant amplitudes (including electroweak penguin contributions).  Sec.~V
summarizes the implications of the deduced amplitudes for other $VP$ processes,
tabulating the 90\% confidence level (c.l.) ranges of predictions, both with
and without tree-penguin interferences which might shed light on the sign of
$\cos \gamma$ or $\cos \alpha$.

Sec. VI compares the results of our analysis with information about the size
of amplitudes and the sign of tree-penguin interference in $B \to PP$
processes, and notes the prospects for obtaining complementary information
from $B_s$--$\bar B_s$ mixing and $K^+ \to \pi^+ \nu \bar \nu$ decays.
Sec.~VII summarizes our arguments for negative $\cos \gamma$ and $\cos \alpha$,
while an Appendix contains a short dictionary relating our invariant amplitudes
in flavor SU(3) to quantities discussed in the factorization approximation, and
discusses our findings in the context of the factorized-amplitude language.
\bigskip
% \newpage

\centerline{\bf II.  NOTATION AND DESCRIPTION OF PROCESSES}
\bigskip

In Table I we list some $VP$ modes of nonstrange $B$ mesons, their
decomposition in terms of reduced amplitudes, and values or 90\% c.l. upper
limits for their branching ratios. We take amplitudes corresponding to the
quark diagrams \cite{Chau,GHLR} $T$ (tree), $C$ (color-suppressed), $P$
(QCD-penguin), $S$ (additional penguin involving flavor-SU(3)-singlet mesons),
$E$ (exchange), $A$ (annihilation) and $PA$ (penguin annihilation). The last
three amplitudes, in which the spectator quark enters into the decay
Hamiltonian, will be neglected here.  Such contributions may be important in
the presence of rescattering, for which tests exist \cite{resc}.  Electroweak
penguin contributions \cite{EWP} are taken into account using the substitution
\cite{GHLRP} 
$$
T \to t \equiv T + P^C_{EW}~~,~~~
C \to c \equiv C + P_{EW}~~,
$$
\beq \label{eqn:combs}
P \to p \equiv P - {1 \over 3}P^C_{EW}~~,~~~
S \to s \equiv S - {1 \over 3}P_{EW}~~,
\eeq
where $P_{EW}$ and $P^C_{EW}$ are color-favored and color-suppressed
electroweak penguin amplitudes. 

We use the phase conventions of Ref.~\cite{GHLR} for pseudoscalar mesons, the
mixing assumption $\eta = (s \bar s - u \bar u - d \bar d)/\st$ and $\eta' = (u
\bar u + d \bar d + 2 s \bar s)/\sx$, and the corresponding phase conventions
for vector mesons with $\rho = (d \bar d - u \bar u)/\s$, $\omega = (u \bar u +
d \bar d)/\s$, and $\phi = s \bar s$. We denote strangeness-preserving ($\Delta
S = 0$) amplitudes by unprimed letters and strangeness-changing ($|\Delta S| =
1$) amplitudes by primed letters.  The suffix on each amplitude denotes whether
the spectator quark is included in a pseudoscalar ($P$) or vector ($V$) meson. 
(For some additional processes not listed in Table I, see Ref.~\cite{DGR}. No
experimental results have been quoted for those processes.) 

A process of the form $B \to X$, where the charge of $B$ is not specified, will
refer to both $B^+$ and $B^0$.  Unless explicitly stated otherwise, we will
always take the branching ratio for a process to refer to the average of that
process and its charge-conjugate. 
\bigskip

\begin{table} \label{amps}
\caption{$B$ decay modes with contributing amplitudes and experimental
branching ratios or 90\% c.l. upper limits (from Ref.~\protect\cite{CLEOVP}
unless stated otherwise).}
\begin{center}
\begin{tabular}{r c c c} \hline \hline
Decay & Amplitudes &
   \multicolumn{2}{c}{Branching ratio (units of $10^{-6}$)}\\
mode  & & Value ($\sigma$) & Upper limit \\
 \hline
\multicolumn{4}{c}{$B^+$ Decays} \\ \hline
$\rho^+ \pi^0$ & $(-t_P+p_V-p_P-c_V)/\s$   & & 77 (a) \\
$\rho^0 \pi^+$ & $(-t_V+p_P-p_V-c_P)/\s$  & $15 \pm 5 \pm 4~(5.2\sigma)$ & \\
$\omega \pi^+$ & $ (t_V+p_P+p_V+c_P+2s_P)/\s$ & $11.3^{+3.3}_{-2.9} \pm
   1.5~(6.2 \sigma)$ & 17 \\
$\phi \pi^+$   &  $s_P$                   & & 4  \\
$\rho^+ \eta$  & $-(t_P+p_P+p_V+c_V+s_V)/\st$ & $4.3^{+4.3}_{-3.4} \pm
   0.7~(1.3 \sigma)$ & 16 \\
$\rho^+ \eta'$ & $(t_P+p_P+p_V+c_V+4s_V)/\sx$ & & 47 \\
$\rho^+ K^0$   & $p'_V$                   & & 48 (a) \\
$\rho^0 K^+$   & $-(p'_V+t'_V+c'_P)/\s$   & & 22 \\
$\omega K^+$   & $(p'_V+t'_V+c'_P+2s'_P)/\s$ & $3.2^{+2.4}_{-1.9} \pm
   0.8~(2.1 \sigma)$ & 8 \\
$\phi K^+$     & $p'_P+s'_P$              & $1.6^{+1.9}_{-1.2} \pm
   0.2~(1.3 \sigma)$ & 5.9 \\
$K^{*0} \pi^+$ &  $p'_P$                  & & 27 \\
$K^{*+} \pi^0$ & $-(p'_P+t'_P+c'_V)/\s$   & & 99 (a) \\
$\bar K^{*0} K^+$ & $p_P$                  & & 12 \\
$K^{*+} \eta$  & $(p'_V-p'_P-t'_P-c'_V-s'_V)/\st$ & $27.3^{+9.6}_{-8.2} \pm 
   5.0~(4.8 \sigma)$ & \\
$K^{*+} \eta'$ & $(2p'_V+p'_P+t'_P+c'_V+4s'_V)/\sx$ & & 87 \\
 \hline 
\multicolumn{4}{c}{$B^0$ Decays} \\ \hline
$\rho^- \pi^+$ & $-t_V-p_V$   & (b) & \\
$\rho^+ \pi^-$ & $-t_P-p_P$   & (b) & \\
$\rho^0 \pi^0$ & $(p_P+p_V-c_P-c_V)/2$    & & 5.1 \\
$\omega \pi^0$ & $(p_P+p_V+c_P-c_V+2s_P)/2$ & & 5.8 \\
$\phi \pi^0$   & $s_P/\s$                 & & 5.4 \\
$\rho^0 \eta$  & $-(p_P +p_V+c_V-c_P+s_V)/\sx$ & $2.6^{+3.0}_{-2.4} \pm
   0.3~(1.3 \sigma)$ & 11 \\
$\rho^0 \eta'$ & $(p_P+p_V+c_V-c_P+4s_V)/(2 \st)$ & & 23 \\
$\rho^- K^+$   & $-p'_V-t'_V$             & & 25  \\
$\rho^0 K^0$   & $(p'_V-c'_P)/\s$         & & 27 \\
$\omega K^0$   & $(p'_V+c'_P+2s'_P)/\s$   & $10.0^{+5.4}_{-4.2} \pm
   1.5~(3.9 \sigma)$ & 21 \\ 
$\phi K^0$     & $p'_P+s'_P$              & $10.7^{+7.8}_{-5.7} \pm 1.1~(2.6
   \sigma)$ & 28 \\
$K^{*+} \pi^-$ & $-p'_P-t'_P$            & $22^{+8+4}_{-6-5}~(5.9 \sigma)$ & \\
$K^{*0} \pi^0$ & $(p'_P-c'_V)/\s$         & & 4.2 \\
$K^{*0} \eta$  & $(p'_V-p'_P-c'_V-s'_V)/\st$ & $13.8^{+5.5}_{-4.4}\pm 1.7~(5.1
   \sigma)$ & \\
$K^{*0} \eta'$ & $(2p'_V+p'_P+c'_V+4s'_V)/\sx$ & & 20 \\
$K^{*+} K^-$   &  (c)                     & & 6 \\ \hline \hline
\end{tabular}
\end{center}
\leftline{(a) From Ref.~\protect\cite{oldVP}.}
\leftline{(b) Sum of $\rho^- \pi^+$ and $\rho^+ \pi^-$ is
$35^{+11}_{-10}\pm 5~(5.6 \sigma)$.}
\leftline{(c) Contributions of order $f_B/m_B$ or rescattering effects.}
\end{table}

\centerline{\bf III.  PATTERNS OF DOMINANT AMPLITUDES}
\bigskip

Using the observed branching ratios quoted in Table I, we can identify reduced
amplitudes for which there exists evidence. We shall assume the lifetimes of
$B^0$ and $B^+$ are equal (valid to a few percent), and shall quote squares of
amplitudes in units of branching ratios with a common factor of $10^{-6}$. 
Thus, an amplitude of 1 will correspond to a branching ratio of $10^{-6}$.  We
make qualitative estimates here, reserving more precise ones for Sec.~IV. 
\bigskip

\leftline{\bf A.  Evidence for $t_V$ and $t_P$}
\bigskip

The decays $B^+ \to \rho^0 \pi^+$ and $B^+ \to \omega \pi^+$, with branching
ratios of order $10^{-5}$, are expected to be dominated by the tree amplitude
$t_V$. The penguin amplitudes $p_P$ and $p_V$ are expected to be smaller than
the corresponding strangeness-changing amplitudes $p'_P$ and $p'_V$ by roughly
a factor of $|V_{td}/V_{ts}| \simeq \lambda$, where $\lambda \simeq 0.22$ is
the parameter introduced by Wolfenstein \cite{WP} to describe the hierarchy of
CKM matrix elements.  The amplitudes $p'_P$ and $p'_V$, as will be seen below,
dominate processes whose branching ratios are of order $10^{-5}$, so one can
conclude that $(|p_V|,|p_P|) = {\cal O}(\lambda) \times |t_V|$.  The amplitude
$s_P$, which contributes to $B^+ \to \omega \pi^+$ but not to $B^+ \to \rho^0
\pi^+$, involves an $\omega$ connected to the rest of the diagram by at least
three gluons.  It is expected to be small by virtue of the Okubo-Zweig-Iizuka
(OZI) rule, which holds well for vector mesons.  Both it and the related
amplitude $s'_P$ will be neglected, except for an electroweak penguin
contribution to $s'_P$ which will be studied in Sec.~IV A.  However, $S'_V$
need not be small, since it involves flavor-singlet couplings to pseudoscalar
mesons for which the OZI rule is less well satisfied.  A related amplitude $S'$
was found important \cite{DGReta} for the decays $B \to K \eta'$.  We shall
take account of $S'_V$ when describing the decays $B \to K^* \eta$ and $B \to
K^* \eta'$. The corresponding $\Delta S=0$ amplitude $S_V$ plays a role in $B
\to \rho \eta$ and $B \to \rho \eta'$. 

We conclude from Table I that $|t_V|^2/2 = {\cal O}(12)$ (to be multiplied, as
mentioned above, by $10^{-6}$ to obtain the corresponding branching ratios for
$B^+ \to \rho^0 \pi^+$ and $B^+ \to \omega \pi^+$).  The possibility that these
two branching ratios differ from one another (as suggested, for example, in
Ref.~\cite{HY}) will be noted in Sec.~IV when we come to discuss interfering
subdominant amplitudes. 

The decay $B^0 \to \rho^- \pi^+$ is also expected to be dominated by $t_V$. In
the absence of separate branching ratios for this process and for $B^0 \to
\rho^+ \pi^-$ (which we expect to be dominated by $t_P$, as will be seen
presently), all that is measured is the sum 
\beq
{\cal B}(B^0 \to \rho^- \pi^+) + {\cal B}(B^0 \to \rho^+ \pi^-)
= (35^{+11}_{-10} \pm 5) \times 10^{-6}~~~,
\eeq
which is consistent with the contribution from $|t_V|^2$ alone, but permits an
additional $|t_P|^2$ contribution.  Identification of the flavor of the
decaying neutral $B$ will distinguish these two decay modes from one another. 
Meanwhile, we anticipate indirect arguments in the next Section that bracket
$|t_P|^2$ between 6.1 and 29. 
\bigskip

\leftline{\bf B.  Evidence for $p'_P$}
\bigskip

The decay $B^0 \to K^{*+} \pi^-$ has been seen at the $5.9 \sigma$ level with a
branching ratio of $(22^{+8+4}_{-6-5}) \times 10^{-6}$.  It is expected to be
dominated by the penguin amplitude $p'_P$.  The other contributing amplitude,
$t'_P$, should be of order $|V_{us}/V_{ud}| = \lambda/(1-\lambda^2/2)$ times
$t_P$, and since $|t_P|^2 < 29$ we expect that $|t'_P| < 1.22$.

As we shall see in Sec.~IV B, it is likely that $B^0 \to K^{*+} \pi^-$ receives
some enhancement from constructive interference between $p'_P$ and other
amplitudes, notably $t'_P$ and an electroweak penguin contribution.  The
branching ratio for $B^+ \to \phi K^+$, also dominated by $p'_P$, is less than
$5.9 \times 10^{-6}$, while that for $B^0 \to \phi K^0$ is $(10.7^{+7.8}_{-5.7}
\pm 1.1) \times 10^{-6}$. These constraints suggest that $|p'_P|^2 \siml {\cal
O}(10)$. For comparison, values of $|p'|^2$ around 18 characterize $B \to K
\pi$ decays, as will be shown in Sec.~VI.  (See also \cite{BPP}.) 
\bigskip

\leftline{\bf C.  Evidence for $p'_V$}
\bigskip

The amplitude $p'_V$ is expected to be very small in all
factorization-dependent calculations of which we are aware
\cite{HHY,AKL,HY,Chau,VPr}, except for one \cite{HSW} which appeared after the
submission of this paper. In those calculations, except the one, all the decays
$B \to (\omega,\rho)K$ are highly suppressed. (See, however, \cite{Ciu}). 

In our earlier analysis of $B \to VP$ decays \cite{DGR}, we concluded that
$p'_V \ne 0$ on the basis of the branching ratio ${\cal B}(B^+ \to \omega K^+)
= (15 \pm 7 \pm 3) \times 10^{-6}$ reported by the CLEO Collaboration
\cite{OldVP}.  With a larger data sample and a revised analysis, the evidence
for this mode has now become considerably weaker, with a 90\% c.l. upper limit
of $8 \times 10^{-6}$ for the branching ratio \cite{CLEOVP}.  However, the
decay $B^0 \to \omega K^0$ indicates that $p'_V \ne 0$, with ${\cal B}(B^0 \to
\omega K^0) = (10.0^{+5.4}_{-4.2} \pm 1.5) \times 10^{-6}$, a $3.9 \sigma$
signal.  As we shall see below, it is likely that $B^+ \to \omega K^+$ receives
destructively interfering contributions from smaller $t'_V$  and electroweak
penguin amplitudes. 

Some time ago the CLEO Collaboration reported evidence for $B^{+,0} \to K^{+,0}
\eta'$ \cite{Oldetap} at a substantial rate, supported by the data sample now
analyzed \cite{CLEOeta}:  ${\cal B}(B^+ \to K^+ \eta') = (80^{+10}_{-9} \pm 8)
\times 10^{-6}$, ${\cal B}(B^0 \to K^0 \eta') = (88^{+18}_{-16} \pm 9) \times
10^{-6}$.  Our interpretation of this result \cite{DGReta} relies in part on a
large flavor-singlet amplitude $s'$ which was proposed previously to the
discovery of these processes \cite{DGRsing}.  However, an additional feature
contributing to $\eta'$ production through the penguin amplitude $p'$ is
constructive interference between contributions from nonstrange and strange
quarks in the $\eta'$, as originally proposed by Lipkin \cite{HJLeta}. 

The CLEO Collaboration has now found evidence for $B^{+,0} \to K^{*(+,0)} \eta$
\cite{CLEOeta}.  Lipkin argues that the same mechanism favoring $B \to K \eta'$
(with $B \to K \eta$ not yet detected) should favor $B \to K^* \eta$, with
constructive interference between $p'_P$ and $p'_V$ in those decays and
destructive interference in $B \to K^* \eta'$.  His argument is equivalent to
the assumption $p'_V = - p'_P$.  A similar relation applies to certain
contributions to charmed particle decays \cite{DD}. 

We shall find that in contrast to the model-dependent predictions based on
factorization for penguin amplitudes, but in accord with Lipkin's suggestion,
there is some evidence for the relation $p'_V \simeq -p'_P$, with $|p'_P|^2
\simeq |p'_V|^2$. The conclusion $p'_V \simeq - p'_P$ then entails $p_V \simeq
- p_P$, causing the penguin contributions to many amplitudes in Table I to
cancel one another. 
\bigskip

\centerline{\bf IV.  INCLUSION OF SUBDOMINANT AMPLITUDES}
\bigskip

We summarize in Table II the evidence for interfering amplitudes to be
discussed in the present Section.  While much of the evidence is not yet
statistically compelling, increased statistics that will be available from the
CLEO III, BaBar, and Belle detectors should be able to confirm or refute the
trends that are suggested by present data.  By taking account of subdominant
contributions, we can arrive at more accurate estimates of those amplitudes
mentioned in Section III. 
\newpage

\leftline{\bf A.  Contributions of electroweak penguins}
\bigskip

\begin{table} \label{ints}
\caption{Patterns in $B \to VP$ data suggesting interference between
dominant and subdominant amplitudes.  All inequalities are those following
from small final-state phases and either $\cos \gamma < 0$ ($|\Delta S| = 1$
processes) or $\cos \alpha > 0$ ($\Delta S = 0$ processes).}
\begin{center}
\begin{tabular}{c c c} \hline \hline
Subdominant & Interfering & Consequence \\
amplitude   &    with     &             \\ \hline
$t'_P$      & $p'_P$      & $\Gamma(K^{*+} \pi^-) > \Gamma(\phi K^{+,0})$ \\
$t'_P$      & $p'_P-p'_V$ & $\Gamma(K^{*+} \eta) > \Gamma(K^{*0} \eta)$ \\
$p_P$       & $t_V$       & $\Gamma(\rho^0 \pi^+) > \Gamma(\omega \pi^+)$ \\
$t'_V$      & $p'_V$      & $\Gamma(\omega K^0) > \Gamma(\omega K^+)$ \\
\hline \hline
\end{tabular}
\end{center}
\end{table}

We shall neglect the color-suppressed contributions $C'_{P,V}$ in
strangeness-changing processes since they are highly suppressed relative to the
dominant penguin amplitudes. The strangeness-conserving terms $C_{P,V}$ are
subdominant to the color-allowed terms $T_{P,V}$. In $B^+$ decays these
amplitudes occur in two specific combinations, $T_P + C_V$ and $T_V + C_P$, to
which we will refer for simplicity as $t_P$ and $t_V$, respectively. Since in
neutral $B$ decays $T_P$ and $T_V$ occur unaccompanied by $C_V$ and $C_P$, this
introduces a small uncertainty in estimating $t_P$ and $t_V$ from the measured
$\Delta S=0$ $B^0$ decay rates. We will use these measurements only to obtain
an upper limit on $|t_P|$. 

The amplitudes $c' = C' + P'_{EW}$ and $s' = S' - (1/3)P'_{EW}$
contain color-favored electroweak penguin amplitudes, which must be taken into
account \cite{EWP}. We employ calculations based on the factorization
hypothesis \cite{AKL,RFDH} to estimate their importance.  This assumption can
be tested once data become precise enough to specify the electroweak penguins
directly. 

We consider only strangeness-changing electroweak penguins, since they are
approximately $1/\lambda^2 \simeq 20$ times as large in amplitude as
strangeness-preserving ones.  Furthermore, we consider only color-favored
amplitudes, which appear only in the case of neutral-meson production. These
are associated, through Eqs.~(1), with the amplitudes $c'_{P,V}$ and
$s'_{P,V}$, and the coefficients of their reduced matrix elements can be
calculated either from Eqs.~(1) and Table I or directly via the wave functions
of the corresponding neutral mesons and quark charges \cite{GHLRP}. 

We need estimates of the electroweak penguin amplitudes ${P'}^P_{EW}$, in which
the spectator quark is incorporated into a pseudoscalar meson, and
${P'}^V_{EW}$, in which the spectator quark is incorporated into a vector
meson.  Strangeness-changing $B$ decays with production of the vector mesons
$\rho^0$, $\omega$, and $\phi$ involve ${P'}^P_{EW}$, while those with
production of $\pi^0$, $\eta$, and $\eta'$ involve ${P'}^V_{EW}$. 

The amplitude ${P'}^P_{EW}$ has been estimated \cite{RFDH} to result in a 30\%
reduction in the predicted rate for $B^+ \to \phi K^+$ in comparison with the
contribution from the gluonic penguin $p'_P$ alone.  From Eqs.~(1) and Table I,
\beq
|A(B^+ \to \phi K^+)|^2 = |p'_P - \frac{1}{3}{P'}^P_{EW}|^2 = 0.7|p'_P|^2~~~,
\eeq
implying ${P'}^P_{EW} \simeq  p'_P/2$ for zero relative strong phase between
the electroweak and gluonic penguins.  (The weak phases of the two are the
same.)  We shall assume this to be the case in what follows. The calculation of
Ref.~\cite{AKL} gives, for $N_c = 2$, an electroweak penguin ${P'}^P_{EW}$
about 2/3 of that in Ref.~\cite{RFDH}, reaching the latter estimate for a 
higher value of $N_c$. (Here $N_c$ is the effective number of
quark colors in a $1/N_c$ expansion).

A more general fit to amplitudes could leave the relative strong phase between
electroweak and gluonic penguin amplitudes as a free parameter.  The relative
strength of electroweak and gluonic penguin amplitudes can be tested by
relating $B \to \phi K$ (which involves both) to $B^+ \to K^{*0} \pi^+$ (which
involves just $p'_P$). With the magnitude of ${P'}^P_{EW}$ estimated in
Ref.~\cite{RFDH}, the corresponding nonstrange contribution $P^P_{EW}$ results
in a predicted branching ratio ${\cal B}(B^+ \to \phi \pi^+) \simeq 10^{-8}$. 
This is completely consistent with our ansatz ${P'}^P_{EW} = p'_P/2$ if
$|p'_P|^2 \simeq 10$ as noted in Sec.~III and if $|P^P_{EW}| \simeq \lambda
|{P'}^P_{EW}| \simeq |p'_P/10|$. 

The decay $B^+ \to \rho^+ K^0$ is pure $p'_V$, while $B^0 \to \rho^0 K^0$
involves interference between $p'_V$ and ${P'}^P_{EW}$.  As follows from the
expectation that $p'_V \simeq - p'_P$ (see Subsection IV C), this interference
is expected to be constructive, resulting in more than a two-fold enhancement
of ${\cal B}(B^0 \to \rho^0 K^0)$ relative to the prediction in the absence of
the electroweak penguin.  The comparison of rates for these two processes thus
is a way to learn ${P'}^P_{EW}/p'_V$.  Assuming that these amplitudes are
relatively real, we predict that 
\beq
\frac{2 \Gamma(\rho^0 K^0)}{\Gamma(\rho^+ K^0)} = \left(
1 - \frac{{P'}^P_{EW}}{p'_V} \right)^2 > 1~~~.
\eeq

The amplitude ${P'}^V_{EW}$ is estimated in \cite{AKL} to contribute to $B^+
\to K^{*+} \pi^0$ in such a way that ${P'}^V_{EW}/p'_P$ = (0.28--0.35,
0.23--0.29, 0.14--0.20) for $N_c = (2,~3,~\infty)$.  We shall assume the
nominal value ${P'}^V_{EW}/p'_P = 1/4$. 

The relative phase expected between ${P'}^V_{EW}$ and $p'_P$ in
Refs.~\cite{RFDH} is consistent with that in \cite{AKL}.  Both find destructive
interference in $B^0 \to K^{*0} \pi^0$. By comparing the rates for $B^+ \to
K^{*0} \pi^+$ (pure $p'_P$) and $B^0 \to K^{*0} \pi^0$ (with
gluonic-electroweak penguin interference), one should be able to deduce
${P'}^V_{EW}/p'_P$ from experiment. Assuming that the two amplitudes are
relatively real, we predict that 
\beq
\frac{2 \Gamma(K^{*0} \pi^0)}{\Gamma(K^{*0} \pi^+)} = \left(
1 - \frac{{P'}^V_{EW}}{p'_P} \right)^2 < 1~~~.
\eeq

We summarize the expectations for EWP contributions in Table III.  We quote
estimates of ratios with respect to gluonic penguin amplitudes since they are
probably more reliable than those of absolute magnitudes.  We will see that the
contributions in Table III lead to a self-consistent picture.  Eventually it
will be possible to determine the electroweak penguin contributions themselves
from the data.  Predicted inequalities based on electroweak penguin
contributions are summarized in Table IV. 

\begin{table} \label{EWPs}
\caption{Color-favored electroweak penguin contributions to $|\Delta S| = 1$
$B$ decays.}
\begin{center}
\begin{tabular}{|c c r|c c c|} \hline \hline
Decay & \multicolumn{2}{c|}{${P'}^P_{EW}$} & Decay & \multicolumn{2}{c|}
{${P'}^V_{EW}$} \\
Mode  & Coeff. & Contrib. & Mode & Coeff. & Contrib. \\ \hline
$\rho^0 K$ & $-\frac{1}{\s}$ & $-0.35p'_P$ & $K^* \pi^0$ & $-\frac{1}{\s}$ &
 $-0.18 p'_P$ \\
$\omega K$ & $\frac{1}{3\s}$ & $0.12p'_P$ & $K^* \eta$ & $-\frac{2}{3\st}$ &
 $-0.10 p'_P$ \\
$\phi K$   & $-\frac{1}{3}$ & $-0.17p'_P$ & $K^* \eta'$ & $-\frac{1}{3\sx}$ &
 $-0.03 p'_P$ \\ \hline \hline
\end{tabular}
\end{center}
\end{table}
\bigskip

\begin{table} \label{EWPin}
\caption{Predicted inequalities due to interference between electroweak
and gluonic penguin amplitudes.}
\begin{center}
\begin{tabular}{c c c} \hline \hline
Subdominant & Interfering & Consequence \\
amplitude   &   with      & \\ \hline
${P'}^P_{EW}$ & $p'_P$ & $\Gamma(\phi K^+) = \Gamma(\phi K^0) < \Gamma(K^{*0}
\pi^+)$ \\
${P'}^P_{EW}$ & $p'_V$ & $2 \Gamma(\omega K^0) <
  \Gamma(\rho^+ K^0) < 2 \Gamma(\rho^0 K^0)$ \\
${P'}^V_{EW}$ & $p'_P$ & $\Gamma(K^{*0} \pi^+) > 2 \Gamma(K^{*0} \pi^0)$ \\
\hline \hline
\end{tabular}
\end{center}
\end{table}

\leftline{\bf B.  Evidence for $t'_P$--$p'_P$ interference from $B^0 \to
K^{*+} \pi^-$ and $B \to \phi K$}
\bigskip

We present in this subsection the main evidence for negative $\cos \gamma$,
through the enhancement of the decay $B^0 \to K^{*+} \pi^-$ relative to its
contribution from the penguin amplitude alone. Including the contribution of
the electroweak penguin amplitude, we noted in the previous subsection that
\cite{RFDH} $|A(B \to \phi K)|^2 = 0.7|p'_P|^2$. Using the experimental upper
limit \cite{CLEOVP} ${\cal B}(B^+ \to \phi K^+) < 5.9 \times 10^{-6}$, we then
estimate $|p'_P|^2 < 8.4$ or $|p'_P| < 2.90$ (90\% c.l. as usual).  The
branching ratio for $B^0 \to \phi K^0$ is consistent with this bound. 

On the other hand, ${\cal B}(B^0 \to K^{*+} \pi^-) > 12 \times 10^{-6}$ at 90\%
c.l., implying $\overline{|p'_P + t'_P|^2} > 12$ for the charge-averaged rate.
We also have the constraint $|t'_P| < 1.22$ which was mentioned in Sec.~III B.
The weak phase of $p'_P$ is $\pi$ and the weak phase of $t'_P$ is $\gamma$.  We
temporarily relax our assumption of vanishing strong phases, assuming only that
the relative strong phase $\delta$ between $p'_P$ and $t'_P$ satisfies $\cos
\delta >0$.  This assumption appears reasonable on the basis of both
perturbative \cite{BSS,BBNS} and statistical \cite{SW} estimates.  We then find
that the equations 
$$
|p'_P|< 2.90~~,~~~|t'_P| < 1.22~~,~~~
$$
\beq
\overline{|p'_P + t'_P|^2} = |p'_P|^2 + |t'_P|^2 - 2|p'_P t'_P| \cos \gamma
 \cos \delta > 12
\eeq
have a solution only for the range 
\beq\label{t'P}
2.25 < |p'_P| < 2.90~~,~~~
0.56 < |t'_P| < 1.22~~,~~~
107^\circ < \gamma \le 180^\circ~~~,
\eeq
where the value $\gamma = 107^\circ$ corresponds to the maximum values $|p'_P|
= 2.90$, $|t'_P| = 1.22$, $\cos \delta = 1$. Thus, constructive interference
between $t'_P$ and $p'_P$, corresponding to $\cos \gamma < 0$, is required. 
The value $\gamma = 107^\circ$ is marginally consistent with the range
specified in Ref.~\cite{Falk}, but far outside the aggressive limits quoted in
Ref.~\cite{Parodi}. 
\bigskip

\leftline{\bf C.  Information on $p'_V$ and $S'_V$ from $B \to (\omega K,
K^* \eta, K^* \eta')$}
\bigskip

We define the parameters $\kappa$ and $\mu$ by $S'_V = \kappa p'_P$, $p'_V = -
\mu p'_P$.  We assume $\kappa$ and $\mu$ are real, thereby neglecting a
possible strong phase.  A set of constraints on $\kappa$ and $\mu$ will be
deduced from bounds on the magnitudes of the following amplitudes: 
\beq
A(B^0 \to \omega K^0) = [p'_V + \frac{1}{3}{P'}^P_{EW}]/\s 
  = (-\mu + \frac{1}{6}) p'_P/\s~~~,
\eeq
\beq
A(B^+ \to K^{*+} \eta) = [p'_V - p'_P - t'_P -\frac{2}{3}{P'}^V_{EW} - S'_V]
/\st = -[(\mu + \kappa + \frac{7}{6})p'_P + t'_P]/\st~~~,
\eeq
\beq
A(B^0 \to K^{*0} \eta) = [p'_V - p'_P - \frac{2}{3}{P'}^V_{EW} - S'_V]/\st
= -(\mu + \kappa + \frac{7}{6})p'_P/\st~~~,
\eeq
\beq
A(B^0 \to K^{*0} \eta') = [2p'_V + p'_P - \frac{1}{3}{P'}^V_{EW} + 4 S'_V]/\sx
= (-2 \mu + 4 \kappa + \frac{11}{12})p'_P)\sx~~~.
\eeq
The 90\% c.l. bounds we use, based on the data in Table I, are
$$
4.3 < |A(\omega K^0)|^2 < 17.2~~,~~~15.0 < |A(K^{*+} \eta)|^2~~~,
$$
\beq
|A(K^{*0} \eta)|^2 < 21.2~~,~~~|A(K^{*0} \eta')|^2 < 20~~~.
\eeq
Other experimental bounds turn out to give weaker conditions on $\kappa$ and
$\mu$.  Taking account of the maximum magnitude $|t'_P| < 1.22$, these bounds
lead to the conditions 
$$
2.93 < \left| \mu-\frac{1}{6} \right| |p'_P| < 5.86~~,~~~
5.49 < \left| \mu + \kappa + \frac{7}{6} \right| |p'_P| < 7.97~~~,
$$
\beq \label{eqn:cons}
\left| 4 \kappa - 2 \mu + \frac{11}{12} \right| |p'_P| < 10.95~~~.
\eeq
With $|p'_P| = 2.90$, these relations then have a solution within the
trapezoidal region bounded by the points $(\kappa,\mu) = (-0.45,1.18)$,
$(0.40,1.18)$, $(-0.26,1.84)$, and $(-0.54,1.26)$.  With $|p'_P| = 2.25$ the
corresponding region is bounded by $(-0.20,1.47)$, $(0.91,1.47)$,
$(-0.17,2.55)$, and $(-0.54,1.81)$.  Negative values of $\mu$ violate the last
constraint in Eq.~(\ref{eqn:cons}), leading to unacceptably high values of the
branching ratio for $B^0 \to K^{*0} \eta'$. 

The bound $\mu > 1.18$ is very close to the situation in Ref.~\cite{HJLeta},
where $p'_V = -p'_P$, and very far from the model calculations in which $|p'_V|
\ll |p'_P|$.  Thus, we will assume in what follows that $p'_V = - p'_P$ or $\mu
= 1$, and will tolerate a small discrepancy ($\sim 1.6 \sigma$) with respect to
the observed value of ${\cal B}(B^0 \to \omega K^0)$. 

For $\mu = 1$ and $|p'_P| = 2.90$, the value of $\kappa = S'_V/p'_P$ is bounded
between $-0.28$ and 0.58 by the second of Eqs.~(13). The corresponding range
for $|p'_P| = 2.25$ is $0.27 < \kappa < 1.38$.  We shall take $\kappa = 0.58$. 
In Sec.~VI A we shall show that a similar ratio is characteristic of $S'/P'$ in
$B \to PP$ decays.  As in Ref.~\cite{DGReta}, there is still some uncertainty
in the contribution of the flavor-singlet amplitude which must be resolved with
the help of better data on decays involving $\eta$ and $\eta'$. 

With these parameters, with $|p'_p| < 2.90$, and with maximally constructive
interference, one finds ${\cal B}(B^+ \to K^{*+} \eta) < 28 \times 10^{-6}$. 
With no interference, the maximum branching ratio is about $22 \times 10^{-6}$.
Thus, in order to demonstrate convincing interference between $t'_P$ and the
remaining amplitudes in $B^+ \to K^{*+} \eta$, it will probably be necessary to
measure the difference in branching ratios for $B^{(+,0)} \to K^{*(+,0)} \eta$
to about 10\%. 

The large value of $|p'_V| > 2.25$ from $B^0 \to \omega K^0$ stands in sharp
contradiction to most explicit models.  It implies that ${\cal B}(B^+ \to
\rho^+ K^0) > 5 \times 10^{-6}$. In contrast, for example, Ref.~\cite{HY}
predicts ${\cal B}(B^+ \to \rho^+ K^0) = 6 \times 10^{-7}$, nearly an order of
magnitude lower. 

With $p'_V = - p'_P$, $2.25 < |p'_P| < 2.90$, and $S'_V = 0.58 p'_P$, the
branching ratio for $B^0 \to K^{*0} \eta$ is predicted to range between 13 and
$21 \times 10^{-6}$, in satisfactory agreement with the experimental value. 
\bigskip

\leftline{\bf D.  Information on $t_V$ and $t_P$ from $B^+ \to (\rho^0,\omega)
\pi^+$ and $B^0 \to \rho^\mp \pi^\pm$ rates}
\bigskip

As mentioned in Sec.~III A, the dominant amplitude contributing to $B^+ \to
\rho^0 \pi^+$ and $B^+ \to \omega \pi^+$ is $t_V$.  The central values of the
branching ratios quoted in Table I correspond to differences of tens of percent
between these two rates. This trend is what one expects if the dominant
amplitude $t_V$ interferes with $p_P$ constructively in $\rho^0 \pi^+$ and
destructively in $\omega \pi^+$.  From the bounds on $p'_P$ obtained earlier
and the expectation $|p_P| \simeq \lambda |p'_P|$, one finds $0.49 < |p_P| <
0.64$ (90\% c.l.). The relative phase between $p_P$ and $t_V$ should be $\beta
+ \gamma = \pi - \alpha$ in the limit that one neglects up and charmed quark
contributions \cite{BFC} to $p_P$. Thus, for $\cos \alpha > 0$, one will get
constructive interference between $t_V$ and $p_P$ in $\rho^0 \pi^+$ and
destructive interference in $\omega \pi^+$. 

We should mention the importance of the up and charmed quark contributions to
$\bar b \to \bar d$ penguin amplitudes noted in Ref.~\cite{BFC}.  In the limit
in which the charm contribution dominates, one measures $- \cos \gamma$ instead
of $\cos \alpha$.  Since $\gamma = \pi - \alpha - \beta$ and $\beta$ is limited
to a small range around $20^\circ$, the values of $\cos \alpha$ and $- \cos
\gamma$ are not all that different.  However, without an explicit evaluation of
the relative up, charm, and top contributions to the $\bar b \to \bar d$
penguins, the interpretation of tree-penguin interference in $\Delta S = 0$
processes is open to some question \cite{LSS}. 

The coefficient of $p_V$ is of the same sign as that of $t_V$ in $B^+ \to
\rho^0 \pi^+$ and $B^+ \to \omega \pi^+$.  With the conjectured relation $p_V =
- p_P$ mentioned at the end of Section III, one then has 
\beq
A(B^+ \to \rho^0 \pi^+) = (-t_V + 2p_P)/\s~~,~~~
A(B^+ \to \omega \pi^+) = t_V/\s~~~,
\eeq
so that one may use the latter process to estimate $|t_V|^2$, with the result 
\beq
2 |A(B^+ \to \omega \pi^+)|^2 = |t_V|^2 = 22.6^{+7.3}_{-6.5} ~~~,
\eeq
or $3.78 < |t_V| < 5.65$, $ 0.85 <|t'_V| = |V_{us}/V_{ud}||t_V| < 1.28$ at 90\%
c.l. Assuming top-quark dominance of $p_P$, the phase $\alpha$ can then in
principle be deduced from the rate for $B^+ \to \rho^0 \pi^+$.  With present
data (neglecting strong phase differences to display maximal interference
effects), the rate 
\beq
2 |A(B^+ \to \rho^0 \pi^+)|^2 = ||t_V| + 2 e^{i \alpha} |p_P||^2
= 30 \pm 13~~~,
\eeq
still permits all values of $\cos \alpha$.  However, if we were to take the
central value of the $\omega \pi^+$ branching ratio seriously, we would obtain
$B(\rho^0 \pi^+) = (17,12) \times 10^{-6}$ for (fully constructive, no)
interference between $t_V$ and $p_V-p_P = - 2p_P$.  As in the case of the $K^*
\eta$ decays mentioned above, a 10\% measurement of the $\omega \pi^+$ and
$\rho \pi^+$ branching ratios would be needed to begin to shed light on the
expected magnitude of the interference term. 

We now use data on $B^0 \to \rho^\mp \pi^\pm$ to specify the upper bound
(mentioned in Sec.~III A) on $|t_P|$.  We use the fact that in $|A(\rho^-
\pi^+)|^2 = |t_V + p_V|^2$ and $|A(\rho^+ \pi^-)|^2 = |t_P + p_P|^2$, the
relations $p_V \simeq - p_P$ and $t_V \simeq t_P$ cause the interference terms
to cancel approximately, so that the sum of the rates yields 
\beq
|t_V|^2 + |t_P|^2 + |p_V|^2 + |p_P|^2 \simeq 35^{+12.1}_{-11.2}~~~.
\eeq
Taking $0.49 < |p_V| \simeq |p_P| < 0.64$ and $|t_V|^2$ as specified above, we
find $|t_P|^2 < 29$ at 90\% c.l.  The corresponding upper bound on $|t'_P|$ is
1.22, as mentioned in Sec.~III B. The lower bound on $|t'_P|$ implies $|t_P| >
2.48$, $|t_P|^2 > 6.1$. 
\bigskip
%\newpage
 
\leftline{\bf E.  Effect of $t'_V$ in $\omega K$ decays}
\bigskip

Taking account of electroweak penguin contributions, the amplitudes for $\omega
K$ decays are dominated by the terms 
\beq
A(B^+ \to \omega K^+) = (p'_V + t'_V + 0.17 p'_P)/\s~~~,
\eeq
\beq
A(B^0 \to \omega K^0) = (p'_V + 0.17 p'_P)/\s~~~.
\eeq
With $p'_V = - p'_P$ (as chosen in our analysis of $K^* (\omega,\eta)$ decays),
we then find
\beq
A(B^+ \to \omega K^+) = (t'_V - 0.83 p'_P)/\s~~,~~~
A(B^0 \to \omega K^0) = - 0.83 p'_P/\s~~~.
\eeq
The last equation implies $1.8 \times 10^{-6} < B(B^0 \to \omega K^0) < 2.9
\times 10^{-6}$ for $2.25 < |p'_P| < 2.90$.  The experimental branching ratio
is compatible with this range at $1.6 \sigma$ (95\% c.l.), as mentioned above.
We note in passing that we predict $\Gamma(B^0 \to \omega K^0) = \Gamma(B^+ \to
\phi K^+)/2 = \Gamma(B^0 \to \phi K^0)/2$.  Since $B^0 \to \omega K^0$ has been
seen while $B \to \phi K$ has not (the $\phi K^0$ signal is only $2.6 \sigma$),
this could be the first test capable of falsifying our assumption that $p'_V =
-p'_P$. 

Since the weak phase of $t'_V$ is $\gamma$ and that of $p'_P$ is $\pi$, we get
destructive interference between $t'_V$ and $p'_P$ in the charge-averaged
$\omega K^\pm$ rate (when $\cos \delta > 0$) for $\cos \gamma < 0$.  Taking
$0.85 < |t'_V| < 1.28$ as suggested above, $|p'_P| = 2.90$, and maximally
destructive interference in $B^+ \to \omega K^+$, we get $B(B^+ \to \omega K^+)
= 10^{-6}$.  Thus, although we predict $p'_V$ to have a considerably larger
magnitude than do many specific models, its effect in $B^+ \to \omega K^+$ can
be considerably diminished by destructive interference with both tree and
electroweak penguin contributions. 
\bigskip

\centerline{\bf V.  SUMMARY OF PREDICTIONS FOR $B \to VP$ DECAYS}
\bigskip

We have deduced ranges for the main amplitudes governing $B \to VP$ decays from
existing data and bounds.
Within these ranges, some of the penguin amplitudes ($p_V$, $S_V$,
$P^{P,V}_{EW}$ and the corresponding $|\Delta S| = 1$ amplitudes) were chosen
to have specific ratios relative to $p_P$ or $p'_P$. 
Our results (at 90\% c.l.) are summarized in Table V.  They imply that many of
the processes listed in Table I should be observable at branching ratio levels
of a few parts in $10^{6}$.  These are summarized in Tables VI and VII.  Some
were already noted in the previous Sections.  In addition to predictions for
the sign of interference with $\cos \gamma < 0$ and $\cos \alpha > 0$, shown by
numbers in parentheses and mentioned in the text, predictions for the opposite
sign of interference with $\cos \gamma > 0$ and $\cos \alpha < 0$ are shown in
brackets. 
\bigskip

\begin{table}
 \caption{Summary of 90\% c.l. bounds or assumed values for amplitudes
contributing to $B \to VP$ decays.} 
\begin{center}
\begin{tabular}{c c c c} \hline \hline
\multicolumn{2}{c}{$\Delta S = 0$} & \multicolumn{2}{c}{$|\Delta S| = 1$}
 \\ \hline
Amplitude & Range & Amplitude & Range \\ \hline
$t_V$ & 3.78--5.65 & $t'_V$  & 0.85--1.28 \\
$t_P$ & 2.48--5.39 & $t'_P$  & 0.56--1.22 \\
$p_P$ & 0.49--0.64 & $p'_P$  & 2.25--2.90 \\
$p_V$ &   $-p_P$   & $p'_V$  &  $-p'_P$   \\
$S_V$ &  $0.58p_P$   & $S'_V$  & $0.58p'_P$ \\
$P^P_{EW}$ & $p_P/2$ & ${P'}^P_{EW}$ & $p'_P/2$ \\
$P^V_{EW}$ & $p_P/4$ & ${P'}^V_{EW} $& $p'_P/4$ \\
\hline \hline
\end{tabular}
\end{center}
\end{table}

\leftline{\bf A.  $\rho \pi$ decays}
\bigskip

The decay $B^+ \to \rho^+ \pi^0$ should be dominated by $|t_P|$, whose square
we estimated to be between 6.1 and 29 (in our usual units), leading to a
prediction ${\cal B}(B^+ \to \rho^+ \pi^0)$ between 3.6 and $15 \times 10^{-6}$
in the absence of tree-penguin interference. If interference between $t_P$ and
$p_P-p_V \simeq 2 p_P$ occurs, it is likely to be destructive for $\cos \alpha
>0$, and could reduce the lower bound to less than $10^{-6}$.  In $B^0 \to
\rho^- \pi^+$ we expect 
\beq
A(B^0 \to \rho^- \pi^+) = -(t_V + p_V) \simeq -(t_V - p_P)~~~,
\eeq
with $B(B^0 \to \rho^- \pi^+)$ ranging from $15 \times 10^{-6}$ (for the
smallest acceptable $|t_V|$ and no interference) to $40 \times 10^{-6}$ (for
the largest acceptable $|t_V|$ and constructive interference, as expected if
$\cos \alpha > 0$).  With $A(B^0 \to \rho^+ \pi^-) = -(t_P + p_P)$ we expect
$B(B^0 \to \rho^+ \pi^-)$ ranging from 6.3 to $29 \times 10^{-6}$ (for no
interference) or as low as $3.3 \times 10^{-6}$ (with destructive interference,
as expected if $\cos \alpha > 0)$. 

\begin{table} \label{plus}
\caption{Predicted 90\% c.l. ranges for $B^+ \to VP$ branching ratios in units
of $10^{-6}$.  Values are quoted for no tree-penguin interference.  Those in
parentheses denote possible values when such interference is present with $-1 <
\cos \gamma < 0$ or $0 < \cos \alpha < 1$, while those in brackets denote
corresponding possible values with $0 < \cos \gamma < 1$ or $-1 < \cos \alpha <
0$.  Charge-averaged branching ratios are understood here.} 
\begin{center}
\begin{tabular}{c c} \\ \hline \hline
Mode & Predicted b.r. \\ \hline
$\rho^+ \pi^0$ & 3.6~(0.7)--15~[22] \\
$\rho^0 \pi^+$ & 7.8~[3.0]--17~(24) \\
$\omega \pi^+$ & 7.1--16 (a) \\
$\rho^+ \eta$  & 2.1~(1.5)--9.7~[11] \\
$\rho^+ \eta'$ & 1.2~(0.2)--5.2~[7.9] \\
$\rho^+ K^0$   & 5--8.4 \\
$\rho^0 K^+$   & 1.0~(0)--1.9~[3.7] \\
$\omega K^+$   & 2.1~(0.2)--3.8~[6.8] \\
$\phi K^+$     & 3.5--5.9 (b) \\
$K^{*0} \pi^+$ & 5--8.4 \\
$K^{*+} \pi^0$ & 4.1~[1.3]--7.3~(12) \\
$K^{*+} \eta$ & 13~[8.2]--22~(28) \\
$K^{*+} \eta'$ & 1.3~[0.4]--2.4~(3.8) \\ \hline \hline
\end{tabular}
\end{center}
\leftline{\qquad \qquad \qquad (a) Input value was $11.3^{+3.6}_{-3.3}$}
\leftline{\qquad \qquad \qquad (b) Upper bound served as input}
\end{table}

\begin{table} \label{neut}
\caption{Predicted ranges for $B^0 \to VP$ branching ratios.
(See caption to Table VI for details.)}
\begin{center}
\begin{tabular}{c c} \\ \hline \hline
Mode & Predicted b.r. \\ \hline
$\rho^- \pi^+$ & 15~[11]--32~(40) \\
$\rho^+ \pi^-$ & 6.3~(3.3)--29~[37] \\
$\rho^- K^+$   & 5.6~(0.9)--10~[17] \\
$\rho^0 K^0$   & 5.7--9.5 \\
$\omega K^0$   & 1.8--2.9 \\
$\phi K^0$     & 3.5--5.9 \\
$K^{*+} \pi^-$ & 5.4~[1.1]--9.9~(17) (a) \\
$K^{*0} \pi^0$ & 1.4--2.4 \\
$K^{*0} \eta$  & 13--21 (b) \\
$K^{*0} \eta'$ & 1.3--2.4 \\
\hline \hline
\end{tabular}
\end{center}
\leftline{\qquad \qquad \qquad (a) Input value was $22^{+9}_{-8}$}
\leftline{\qquad \qquad \qquad (b) Upper bound served as input}
\end{table}
\bigskip

\leftline{\bf B.  $\rho^+ \eta$ and $\rho^+ \eta'$ decays}
\bigskip

The decays $B^+ \to \rho^+ \eta$ and $B^+ \to \rho^+ \eta'$ should be
characterized by branching ratios ranging from 2 to $11 \times 10^{-6}$ and 0.2
to $8 \times 10^{-6}$, respectively.  Some events of $\rho^+ \eta$ have
already been observed \cite{CLEOVP}, but the significance of the signal is
marginal.  This process may be a useful one to measure the magnitude of $t_P$,
since the (small) penguin contributions $p_V$ and $p_P$ are expected to cancel
one another approximately. 
\bigskip

\leftline{\bf C.  $\rho K$ decays}
\bigskip

The decay $B^+ \to \rho^+ K^0$ is due entirely to the $p'_V$ amplitude.  Thus
we expect ${\cal B}(B^+ \to \rho^+ K^0)$ to range between 5 and $8.4 \times
10^{-6}$.  With $|p_V|^2 \simeq |p'_V|^2/20$ we then expect ${\cal B}(B^+ \to
K^{*+} \bar K^0)$ and ${\cal B}(B^0 \to K^{*0} \bar K^0)$ to range between 0.3
and $0.4 \times 10^{-6}$.  At these low levels it is hard to exclude the
possibility that rescattering effects could feed these channels from other more
abundant ones. 

The decay $B^+ \to \rho^0 K^+$ should have a branching ratio very similar to
that for $B^+ \to \omega K^+$ except for the electroweak penguin contribution.
However, this contribution is expected to be appreciable, further reducing the
magnitude of the amplitude through destructive interference with $p'_V$ and
$t'_V$: 
\beq
A(B^+ \to \rho^0 K^+) = -(p'_V + t'_V + 0.5 p'_P)/\s
= -(t'_V - 0.5 p'_P)/\s~~~,
\eeq
leading to a branching ratio of 1 to $1.9 \times 10^{-6}$ if there is no
interference between $t'_V$ and $p'_P$ and possibly even less if there is
destructive interference as expected for $\cos \gamma <0$. 

The decay $B^0 \to \rho^- K^+$ has the same matrix element as $B^+ \to \rho^0
K^+$ except that it is missing the electroweak penguin contribution and is a
factor of $\s$ larger: 
\beq
A(B^+ \to \rho^- K^+) = -(p'_V + t'_V)~~~.
\eeq
With no interference, the expected range of branching ratios is about 5.6 to
$10 \times 10^{-6}$, while destructive interference for $\cos \gamma < 0$ could
push the branching ratio below $10^{-6}$. 

The decay $B^0 \to \rho^0 K^0$ is expected to exhibit constructive interference
between the gluonic and electroweak penguin amplitudes: 
\beq
A(B^0 \to \rho^0 K^0) = \frac{p'_V-c'_P}{\s} \simeq \frac{p'_V-{P'}^P_{EW}}
{\s} \simeq \frac{-p'_P - 0.5 p'_P}{\s}~~~,
\eeq
leading to a predicted range for the branching ratio between 5.7 and $9.5
\times 10^{-6}$, an order of magnitude greater than that in the model of
Ref.~\cite{HHY}. 
\bigskip

\leftline{\bf D.  $\phi K$ decays}
\bigskip

The anticipated limits on $|p'_P|$ lead to ${\cal B}(B^+ \to \phi K^+)$ between
3.5 and $5.9 \times 10^{-6}$.  Suppression below this range would cast doubt
most likely on our assumption of flavor SU(3), which required the amplitude for
$s \bar s$ production in the final state to be the same as that for $u \bar u$
or $d \bar d$ production.  Another possibility is that electroweak penguin
contributions lead to a stronger suppression than in Eq.~(3).  We expect ${\cal
B}(B^+ \to \phi K^+) = {\cal B}(B^0 \to \phi K^0)$, so a similar range is
expected for $B^0 \to \phi K^0$. 
% \bigskip
\newpage

\leftline{\bf E.  $K^* \pi$ decays}
\bigskip

The decay $B^+ \to K^{*0} \pi^+$ is expected to be dominated by the $p'_P$
amplitude, and thus to have a range of branching ratios between 5 and $8.4
\times 10^{-6}$.  The corresponding $\Delta S = 0$ processes, $B^+ \to \bar
K^{*0} K^+$ and $B^0 \to \bar K^{*0} K^0$, should have branching ratios between
0.3 and $0.4 \times 10^{-6}$. 

The gluonic and electroweak penguin amplitudes are the main contributors to
$B^0 \to K^{*0} \pi^0$, leading to an amplitude $A = (p'_P-{P'}^V_{EW})/\s =
0.75 p'_P/\s$.  With $2.25 < |p'_P| < 2.9$, we then predict $1.4 \times 10^{-6}
< {\cal B}(B^0 \to K^{*0} \pi^0) < 2.4 \times 10^{-6}$, in accord with the
upper limit of $4.2 \times 10^{-6}$ for this branching ratio. 

In the decay $B^+ \to K^{*+} \pi^0$, constructive tree-penguin interference can
occur for $\cos \gamma < 0$.  If no such interference occurs, one expects the
branching ratio to range between 4.1 and $7.3 \times 10^{-6}$, while it can
reach as high as $12 \times 10^{-6}$ if the constructive interference is
present. 
\bigskip

\leftline{\bf F.  $K^* \eta$ and $K^* \eta'$ decays}
\bigskip

The tree-penguin interference is expected to be constructive for $\cos \gamma <
0$ in $B^+ \to K^{*+} \eta$, leading to ${\cal B}(B^+ \to K^{*+} \eta) > {\cal
B}(B^0 \to K^{*0} \eta)$.  For the value of $S'_V$ chosen here, it is also
expected to be constructive for $\cos \gamma < 0$ in $B^+ \to K^{*+} \eta'$,
leading to ${\cal B}(B^+ \to K^{*+} \eta') > {\cal B}(B^0 \to K^{*0} \eta')$. 
The exact magnitude of these branching ratios is very sensitive to $S'_V$. The
large disparity between $K^* \eta$ and $K^* \eta'$ is only possible when $p'_V$
has an appreciable magnitude, comparable to that of $p'_P$, in contrast to the
estimates based on factorization. 
\bigskip
%\newpage

\leftline{\bf G. Other processes with neutral mesons}
\bigskip

The processes $B^0 \to (\rho^0,\omega)+ (\pi^0,\eta,\eta')$ are expected to
have very small rates.  The gluonic penguin contributions $p_P$ and $p_V$ are
predicted to cancel one another, leaving only small electroweak penguin terms.
At this level we cannot exclude the possibility that color-suppressed
amplitudes, neglected here, play a role, so we estimate only that these
branching ratios are all less than $10^{-6}$, and do not quote them in Tables
VI and VII. 
\bigskip

\centerline{\bf VI.  IMPLICATIONS FOR OTHER PROCESSES}
\bigskip

\leftline{\bf A.  $B \to PP$ decays}
\bigskip

The CLEO Collaboration \cite{CLEOPP} has presented evidence for two $B \to \pi
\pi$ modes, four $B \to K \pi$ modes and two $B \to K\eta'$ modes
\cite{CLEOeta}, as shown in Table VIII. Also shown are decompositions into
SU(3)-invariant amplitudes, including explicit contributions of color-favored
electroweak penguins.  (The smaller contributions of color-suppressed
electroweak penguins are omitted.) We make several observations about these
results. 

\begin{table} \label{PP}
\caption{$B \to \pi \pi$, $B \to K \pi$ and $B \to K \eta'$ decay modes, with 
decomposition into invariant amplitudes.}
\begin{center}
\begin{tabular}{c c c} \hline \hline
Decay & Amplitudes & B.r.~(units of $10^{-6}$) \\
Mode  &       & ($\sigma$) \\ \hline
$\pi^+ \pi^-$ & $-(T+P)$ & $4.7^{+1.8}_{-1.5} \pm 0.6~(4.2 \sigma)$ \\
$\pi^+ \pi^0$ & $-(T+C+P_{EW})/\s$ & $5.4^{+2.1}_{-2.0}\pm 1.5~(3.2 \sigma)$ \\
$K^+ \pi^-$   & $-(T'+P')$ & $18.8^{+2.8}_{-2.6} \pm 1.3~(11.7 \sigma)$ \\
$K^+ \pi^0$  & $-(T'+P'+C'+P'_{EW})/\s$ & $12.1^{+3.0+2.1}_{-2.8-1.4}~(6.1
  \sigma)$ \\
$K^0 \pi^+$  & $P'$ & $18.2^{+4.6}_{-4.0} \pm 1.6~(7.6 \sigma)$ \\
$K^0 \pi^0$  & $(P'-C'-P'_{EW})/\s$ & $14.8^{+5.9+2.4}_{-5.1-3.3}~(4.7 \sigma)$
 \\
$K^+ \eta'$ & $(3P'+4S'+T'+C'-(1/3)P'_{EW})/\sx$ & $80^{+10}_{-9}\pm 8~(16.8
\sigma)$ \\ 
$K^0 \eta'$ & $(3P'+4S'+C'-(1/3)P'_{EW})/\sx$ & $88^{+18}_{-16} \pm 9~(11.7
\sigma)$ \\
\hline \hline
\end{tabular}
\end{center}
\end{table}

1.  {\it Sum rule for $K \pi$ decay rates.}  Lipkin \cite{HJLI} has noted
that the $B \to K \pi$ rates  satisfy the relation
\beq
\Gamma(K^+ \pi^-) + \Gamma(K^0 \pi^+) = 2[\Gamma(K^+ \pi^0) + 
\Gamma(K^0 \pi^0)]
\eeq
when dominated by the $P'$ amplitude and expanded to leading order in smaller
amplitudes.  A slightly more general proof of this relation was given in
Ref.~\cite{GRComb}.  With the new experimental values, this relation [in units
of (branching ratio $\times 10^6$)] reads 
\beq
37.0^{+5.8}_{-5.2} = 53.8^{+14.7}_{-13.7}~~~,
\eeq
in satisfactory agreement at the $\sim 1 \sigma$ level. 
 
2.  {\it Complete $P'$ dominance.}  In the limit in which all amplitudes except
$P'$ can be neglected one has $\Gamma(K^+ \pi^-) = 2 \Gamma(K^+ \pi^0) =
\Gamma(K^0 \pi^+) = 2 \Gamma(K^0 \pi^0)$, which is also in satisfactory
agreement with experiment.  Thus, at the moment, there are no indications from
$K \pi$ decays for the amplitudes $T'$, $C'$, or $P'_{EW}$. 

3.  {\it Bound on $\gamma$ from $\Gamma(K^0 \pi^+)/[2\Gamma(K^+ \pi^0)]$.} A
test for interference of subdominant amplitudes with $P'$, taking account of
electroweak penguin effects, can be based on the deviation of the ratio $R^*
\equiv \Gamma(B^+ \to K^0 \pi^+)/[2 \Gamma(B^+ \to K^+ \pi^0)]$ from unity
\cite{NRPL,NRPRL,MN,GPY}.  With present data $R^* = 0.75 \pm 0.28$, not
differing significantly from 1. 

4.  {\it Bound on $\gamma$ from $\Gamma(K^+ \pi^-)/\Gamma(K^0 \pi^+)$.} An
earlier test for interference of $T'$ with $P'$ \cite{FM} becomes useful when
the ratio $R \equiv \Gamma(B^0 \to K^+ \pi^-)/\Gamma(B^+ \to K^0 \pi^+)$ lies
below 1:  $\sin^2 \gamma \le R$.  Since with present data $R = 1.03 \pm 0.31$,
no useful bound results.  It may be possible to combine data on charge
asymmetries \cite{CLEOasy} with information on $R$ or $R^*$ to place bounds on
$\gamma$ \cite{GRgamma,NRPRL,MN,GP,RF}. 

5.  {\it Information from $B \to \pi \pi$ decays.}  We may briefly update
previous analyses (see, e.g., \cite{DGReta,BPP}) of $B \to PP$ using the new
results in Table VIII.  The small branching ratio for $B^0 \to \pi^+ \pi^-)$
will be seen to favor $\cos \alpha \ge 0$, in accord with other recent claims
\cite{CLEOPP,HHY,HSW,CY}, but only at the $1 \sigma$ level. 

We first estimate the amplitude $T$ from the rate for the decay $B^+ \to \pi^+
\pi^0$.  We find from Table VIII that $|T+C|^2/2 = 5.4 \pm 2.5$. We need an
estimate of the small contribution $C$ and assume for present purposes the
validity of the calculation in Ref.~\cite{BBNS} whereby Re$(C/T)= {\cal
O}(0.1)$.  (As in estimates of electroweak penguins, we place more reliance
on ratios of amplitudes than on absolute magnitudes.) We then find $|T| = 3.0
\pm 0.7$, in satisfactory agreement with other estimates \cite{LKG,Wurt}. 

The penguin amplitude $P$ may be estimated from the process $B^+ \to K^0
\pi^+$, which is expected to be pure penguin:  $|P'|^2 = 18.2 \pm 4.6$, or
$|P'| = 4.3 \pm 0.5$.  Then $|P| = \lambda |P'| = 0.94 \pm 0.12$. 

Assuming top-quark dominance of $P$ \cite{BFC}, the weak phase $\alpha$ and the
relative strong phase $\delta$ then are constrained by the charge-averaged
$\pi^+ \pi^-$ branching ratio: 
\beq
|T|^2 + |P|^2 - 2|TP| \cos \alpha \cos \delta = 4.7 \pm 1.8~~~,
\eeq
\beq
0.9 \pm 0.9 = \cos \alpha \cos \delta~~~.
\eeq
Thus, assuming $\cos \delta > 0$, one favors $\cos \alpha > 0$ , but only at
the $\sim 1 \sigma$ level. 

{\it 6.  Ratio of singlet to penguin amplitudes $S'/P'$.}  Taking the dominant
terms in the $B^0 \to K^0 \eta'$ decays (thereby avoiding possible
complications associated with the $T'$ contribution to $B^+ \to K^+ \eta'$), 
\beq
A(B^0 \to K^0 \eta') \simeq \frac{3P'+4S'}{\sx}~~~,
\eeq
neglecting color-suppressed and electroweak penguin terms, we can estimate the
ratio $S'/P'$ in the case of constructive interference (the case considered for
the ratio $S'_V/p'_P$ contributing to $B \to K^* \eta$). We find $S'/P' = 0.6
\pm 0.2$, which is very close to the value taken for $S'_V/p'_P$. 

{\it 7.  Reduced amplitudes for $B \to PP$ and comparison with those for $B \to
VP$.}  We summarize the reduced amplitudes for $B \to PP$ amplitudes found in
the previous paragraphs in Table IX.  (We do not include the electroweak
penguin amplitudes discussed in Refs.~\cite{NRPL}-\cite{GP}.)

\begin{table}
\caption{Summary of amplitudes contributing to $B \to PP$ decays.}
\begin{center}
\begin{tabular}{c c c c} \hline \hline \\
\multicolumn{2}{c}{$\Delta S = 0$} & \multicolumn{2}{c}{$|\Delta S| = 1$}
 \\ \hline
Amplitude & Value & Amplitude & Value \\ \hline
$|T|$ & $3.0 \pm 0.7$ & $|T'|$  & $\lambda |T| f_K/f_\pi$ \\
$|P|$ & $\lambda |P'| = 0.94 \pm 0.12$ & $|P'|$  & $4.3 \pm 0.5$ \\
$S$ &  $\lambda |S'|$ & $|S'|$  & $(0.6 \pm 0.2)|P'|$ (a) \\
\hline \hline
\end{tabular}
\end{center}
\leftline{(a) Assuming constructive interference between $S'$ and $P'$ in
$B \to K \eta'$ decays.}
\end{table}

The value of $|T|$ found here is comparable to those found for $|t_P|$ and
$|t_V|$ in Table V.  One might expect (see, e.g., \cite{JLRFM}) that $|t_P/T|
\simeq f_\rho/f_\pi \simeq \s$, which is also consistent with present data. The
value of $|P'|$ is somewhat larger than our values of $|p'_P|$ and $|p'_V|$,
where the smallness of $|p'_P|$ is dictated in part by the need to accommodate
a small branching ratio for $B \to \phi K$. We have already commented on the
fact that $|S'_V/P'_V| \simeq |S'/P'|$ is consistent with present data. 

\bigskip
%\newpage

\leftline{\bf B.  $B_s - \bar B_s$ mixing}
\bigskip

The comparison of the $B^0 - \bar B^0$ mixing parameter $\Delta m_d = 0.464 \pm
0.018$ ps$^{-1}$ and the $B_s - \bar B_s$ mixing parameter $\Delta m_s > 14.3$
ps$^{-1}$ (95\% c.l.) \cite{Blaylock} provides information on $\cos \gamma$. 
Defining $f_B$ and $f_{B_s}$ as the nonstrange and strange $B$ meson decay
constants and $B_B$ and $B_{B_s}$ as the corresponding vacuum-saturation
factors, we have 
\beq
\sqrt{\frac{\Delta m_s}{\Delta m_d}} = \frac{f_{B_s}\sqrt{m_{B_s} B_{B_s}}}
{f_B \sqrt{m_B B_B}} \left| \frac{V_{ts}}{V_{td}} \right|~~~,
\eeq
leading, with $f_{B_s} \sqrt{B_{B_s}}/[f_B \sqrt{B_B}] < 1.25$ \cite{JLRFM}, to
$|V_{ts}/V_{td}| > 4.32$ and $|V_{td}/A \lambda^3| = |1-\rho - i \eta| < 1.05$.
 Thus only a small region of $\cos \gamma < 0$ is allowed.  If the indication
for negative $\cos \gamma$ from $B^0 \to K^{*+} \pi^-$ mentioned in Sec.~IV is
borne out, one should be at the verge of observing a signal, not just a lower
bound, for $\Delta m_s$. 
\bigskip

\leftline{\bf C.  $K^+ \to \pi^+ \nu \bar \nu$}
\bigskip

A recent update of constraints on $K^+ \to \pi^+ \nu \bar \nu$ has been
published \cite{BB}.  With $\cos \gamma$ constrained to lie very close to 0 on
the basis of our $B \to VP$ analysis combined with the strong limit on $\Delta
m_s$, the branching ratio ${\cal B}(K^+ \to \pi^+ \nu \bar \nu)$ is constrained
to lie not far from $10^{-10}$.  This is consistent with the present status of
Brookhaven Experiment E787 \cite{Red}, but more data are expected. 
\bigskip

\centerline{\bf VII.  CONCLUSIONS}
\bigskip

We have studied the decays $B \to VP$, where $V$ is a vector meson and $P$ is a
pseudoscalar meson, in a flavor-SU(3)-invariant analysis.  Two main conclusions
have emerged. 

First, the pattern of interferences between dominant and subdominant
amplitudes, particularly in the decay $B^0 \to K^{*+} \pi^-$ (discussed in
Sec.~IV B), favors a weak phase $\gamma$ in the second quadrant of the
unitarity triangle plot:  $\cos \gamma <0$.  This conclusion, reached under the
assumption that the tree-penguin strong phase-difference in the above process
is smaller that $90^\circ$, agrees with that reached on the basis of more
model-dependent analyses \cite{CLEOPP,HHY,HY,HSW,CY}. It depends to some extent
on an estimate of the magnitude of electroweak penguin contributions, and on
the assumption of flavor SU(3) in relating the penguin contributions in $B \to
\phi K$ decays to those in $B \to K^* \pi$ decays. When combined with the
constraint associated with $\Delta m_s$, this result favors values of $\gamma$
close to $90^\circ$. 

Ratios of other decay rates, including not only those listed in Table II but
also the ratios $\Gamma(B^0 \to \rho^- K^+)/\Gamma(B^+ \to \rho^+ K^0)$ and
$\Gamma(B^0 \to K^{*+} \pi^-)/\Gamma(B^+ \to K^{*0} \pi^+)$ (the $VP$ analogues
of the Fleischer-Mannel \cite{FM} ratio $R$) can shed light on tree-penguin
interferences, permitting constraints on CKM phases with sufficently accurate
data. 

Second, there seems to be evidence for a penguin amplitude, called $p'_V$ in
our notation, at a much higher level than predicted by specific models. This
amplitude contributes to a number of processes, notably $B \to K^* \eta$ and
$B^+ \to \rho^+ K^0$. 

We have predicted a number of $B \to VP$ decay rates to have branching ratios
of a few parts in $10^6$, as shown in Tables VI and VII. Assumptions made about
the relative importance of gluonic and electroweak penguin contributions can be
tested by measuring the ratios of certain decay rates in $\rho K$ and $K^* \pi$
channels.  In particular, if the electroweak penguin amplitude responsible for
the suppression of $B^+ \to \phi K^+$ is smaller than the value we have used
\cite{RFDH}, as suggested, for example, in \cite{AKL}, the argument for $\cos
\gamma < 0$ becomes somewhat stronger. 

The discovery of $B \to VP$ processes at the predicted levels would be strong
evidence that the hierarchy of amplitudes suggested here should be taken
seriously.  One would then have a model-independent way to anticipate the
strength of a whole host of $B$ decays whose study can shed light on the source
of CP violation.  The advent of an upgraded detector and improved collider at
the Cornell Electron Storage Rings, the debut of B-factories at SLAC and KEK,
and the ability of hadron machines to contribute incisively to B physics, all
make the future study of $B \to VP$ decays a potentially rich area for
research. 
\bigskip

\centerline{\bf ACKNOWLEDGMENTS}
\bigskip

We thank L. Wolfenstein and F. W\"urthwein for discussions.  JLR would like to
acknowledge the gracious hospitality of the Physics Department of The Technion.
This work was supported in part by the United States Department of Energy under
Grant No.~DE FG02 90ER40560 and by the United States -- Israel Binational
Science Foundation under Research Grant Agreement 94-00253. 
\bigskip

\centerline{\bf APPENDIX:  A DICTIONARY OF AMPLITUDES}
\bigskip

In this Appendix we give expressions for graphical amplitudes as obtained in
factorization-based calculations. We use the notation of \cite{AKL}. A common
factor [in the SU(3) limit], $\s G_F m_V (\epsilon\cdot q_P)$, is omitted and
we define $Q_1 \equiv -2m^2_\pi/[(m_b + m_u)(m_u + m_d)]$, $Q_5 \equiv
-2m^2_K/[(m_b + m_d)(m_s + m_d)]$.  We assume isospin conservation and neglect
differences between $Q_1$ and $Q_2 \equiv -2m^2_\pi/[(m_b + m_d)(2m_d)]$ and
between $Q_5$ and $Q_4 \equiv -2m^2_K/[(m_b + m_u)(m_s + m_u)]$.  We find
\bea
P'_P &=& V^*_{tb} V_{ts} f_{K^*} F_1^{B \to \pi}(m^2_{K^*}) a_4~,\\
P'_V &=& V^*_{tb} V_{ts} f_K A_0^{B \to \rho}(m^2_K)(a_4 + a_6 Q_5)~,\\
T'_P &=& - V^*_{ub} V_{us} f_{K^*} F_1^{B \to \pi}(m^2_{K^*})  a_1~,\\
T'_V &=& - V^*_{ub} V_{us}  f_K A_0^{B \to \rho}(m^2_K)  a_1~,\\
S'_P &=& V^*_{tb} V_{ts} f_\omega F_1^{B \to K}(m^2_\omega) (a_3 + a_5)~,\\
S'_V &=& V^*_{tb} V_{ts} f_{\eta,\eta'} A_0^{B \to K^*}(m^2_{\eta,\eta'}) (a_3
- a_5)~,\\ 
P'^P_{EW} &=& (3/2) V^*_{tb} V_{ts} f_\omega  F_1^{B \to K}(m^2_\omega) (a_9 +
a_7)~,\\
P'^V_{EW} &=& (3/2) V^*_{tb} V_{ts} f_\pi A_0^{B \to K^*}(m^2_\pi) (a_9 -
a_7)~,\\ 
T_P &=& - V^*_{ub} V_{ud} f_\rho F_1^{B \to \pi}(m^2_\rho) a_1~,\\
T_V &=& - V^*_{ub} V_{ud} f_\pi A_0^{B \to \rho}(m^2_\pi) a_1~,\\
P_P &=& V^*_{tb} V_{td} f_\rho F_1^{B \to \pi}(m^2_\rho) a_4~,\\
P_V &=& V^*_{tb} V_{td} f_\pi A_0^{B \to \rho}(m^2_\pi) (a_4  + a_6 Q_1)~.
\eea
In Eq.~(36) the constants $f_{\eta,\eta'}$ are defined with the same
normalization as $f_\pi$ in terms of the quark content of the corresponding
mesons. 

The smallness of the amplitude $P'_V$ in many factorized approaches
\cite{HHY,AKL,HY,Chau,VPr} follows from a strong cancellation in $a_4 + a_6
Q_5$ which depends on the value chosen for $m_s$.  (The recent treatment of
\cite{HSW} avoids this cancellation by choosing a small value of $m_s$.) The
relations $P'_V = - P'_P$ and $P_V = - P_P$ are not expected to have any
special significance in the factorized approach. Whereas we neglected the
amplitudes $S'_P$ and $S_P$, using a suppression argument based on the OZI rule
for vector mesons, this property is not exhibited in the factorization approach
except for a fortuitous cancellation at a particular value of $N_c$. 
This sensitivity to $N_c$ is a measure of non-factorizing effects, and is one
of the reasons one must appeal to experiment (as advocated for some amplitudes
in \cite{AKL} and employed more generally in \cite{HSW}, for example) to
determine the $a_i$.
\bigskip

% Journal and other miscellaneous abbreviations for references
\def \ajp#1#2#3{Am. J. Phys. {\bf#1}, #2 (#3)}
\def \apny#1#2#3{Ann. Phys. (N.Y.) {\bf#1}, #2 (#3)}
\def \app#1#2#3{Acta Phys. Polonica {\bf#1}, #2 (#3)}
\def \arnps#1#2#3{Ann. Rev. Nucl. Part. Sci. {\bf#1}, #2 (#3)}
\def \art{and references therein}
\def \cmts#1#2#3{Comments on Nucl. Part. Phys. {\bf#1}, #2 (#3)}
\def \cn{Collaboration}
\def \cp89{{\it CP Violation,} edited by C. Jarlskog (World Scientific,
Singapore, 1989)}
\def \dpfa{{\it The Albuquerque Meeting: DPF 94} (Division of Particles and
Fields Meeting, American Physical Society, Albuquerque, NM, Aug.~2--6, 1994),
ed. by S. Seidel (World Scientific, River Edge, NJ, 1995)}
\def \dpff{{\it The Fermilab Meeting: DPF 92} (Division of Particles and Fields
Meeting, American Physical Society, Batavia, IL., Nov.~11--14, 1992), ed. by
C. H. Albright \ite~(World Scientific, Singapore, 1993)}
\def \efi{Enrico Fermi Institute Report No. EFI}
\def \epjc#1#2#3{Euro.~Phys.~J.~C {\bf #1}, #2 (#3)}
\def \epl#1#2#3{Europhys.~Lett.~{\bf #1}, #2 (#3)}
\def \f79{{\it Proceedings of the 1979 International Symposium on Lepton and
Photon Interactions at High Energies,} Fermilab, August 23-29, 1979, ed. by
T. B. W. Kirk and H. D. I. Abarbanel (Fermi National Accelerator Laboratory,
Batavia, IL, 1979}
\def \hb87{{\it Proceeding of the 1987 International Symposium on Lepton and
Photon Interactions at High Energies,} Hamburg, 1987, ed. by W. Bartel
and R. R\"uckl (Nucl. Phys. B, Proc. Suppl., vol. 3) (North-Holland,
Amsterdam, 1988)}
\def \ib{{\it ibid.}~}
\def \ibj#1#2#3{~{\bf#1}, #2 (#3)}
\def \ichep72{{\it Proceedings of the XVI International Conference on High
Energy Physics}, Chicago and Batavia, Illinois, Sept. 6 -- 13, 1972,
edited by J. D. Jackson, A. Roberts, and R. Donaldson (Fermilab, Batavia,
IL, 1972)}
\def \ijmpa#1#2#3{Int. J. Mod. Phys. A {\bf#1}, #2 (#3)}
\def \jpb#1#2#3{J.~Phys.~B~{\bf#1}, #2 (#3)}
\def \jhep#1#2#3{JHEP {\bf#1}, #2 (#3)}
\def \lkl87{{\it Selected Topics in Electroweak Interactions} (Proceedings of
the Second Lake Louise Institute on New Frontiers in Particle Physics, 15 --
21 February, 1987), edited by J. M. Cameron \ite~(World Scientific, Singapore,
1987)}
\def \ky{{\it Proceedings of the International Symposium on Lepton and
Photon Interactions at High Energy,} Kyoto, Aug.~19-24, 1985, edited by M.
Konuma and K. Takahashi (Kyoto Univ., Kyoto, 1985)}
\def \mpla#1#2#3{Mod. Phys. Lett. A {\bf#1}, #2 (#3)}
\def \nc#1#2#3{Nuovo Cim. {\bf#1}, #2 (#3)}
\def \np#1#2#3{Nucl. Phys. {\bf#1}, #2 (#3)}
\def \pisma#1#2#3#4{Pis'ma Zh. Eksp. Teor. Fiz. {\bf#1}, #2 (#3) [JETP Lett.
{\bf#1}, #4 (#3)]}
\def \pl#1#2#3{Phys. Lett. {\bf#1}, #2 (#3)}
\def \pla#1#2#3{Phys. Lett. A {\bf#1}, #2 (#3)}
\def \plb#1#2#3{Phys. Lett. B {\bf#1}, #2 (#3)}
\def \pr#1#2#3{Phys. Rev. {\bf#1}, #2 (#3)}
\def \prc#1#2#3{Phys. Rev. C {\bf#1}, #2 (#3)}
\def \prd#1#2#3{Phys. Rev. D {\bf#1}, #2 (#3)}
\def \prl#1#2#3{Phys. Rev. Lett. {\bf#1}, #2 (#3)}
\def \prp#1#2#3{Phys. Rep. {\bf#1}, #2 (#3)}
\def \ptp#1#2#3{Prog. Theor. Phys. {\bf#1}, #2 (#3)}
\def \ptwaw{Plenary talk, XXVIII International Conference on High Energy
Physics, Warsaw, July 25--31, 1996}
\def \rmp#1#2#3{Rev. Mod. Phys. {\bf#1}, #2 (#3)}
\def \rp#1{~~~~~\ldots\ldots{\rm rp~}{#1}~~~~~}
\def \si90{25th International Conference on High Energy Physics, Singapore,
Aug. 2-8, 1990}
\def \slc87{{\it Proceedings of the Salt Lake City Meeting} (Division of
Particles and Fields, American Physical Society, Salt Lake City, Utah, 1987),
ed. by C. DeTar and J. S. Ball (World Scientific, Singapore, 1987)}
\def \slac89{{\it Proceedings of the XIVth International Symposium on
Lepton and Photon Interactions,} Stanford, California, 1989, edited by M.
Riordan (World Scientific, Singapore, 1990)}
\def \smass82{{\it Proceedings of the 1982 DPF Summer Study on Elementary
Particle Physics and Future Facilities}, Snowmass, Colorado, edited by R.
Donaldson, R. Gustafson, and F. Paige (World Scientific, Singapore, 1982)}
\def \smass90{{\it Research Directions for the Decade} (Proceedings of the
1990 Summer Study on High Energy Physics, June 25--July 13, Snowmass, Colorado),
edited by E. L. Berger (World Scientific, Singapore, 1992)}
\def \stone{{\it $B$ Decays} (Revised 2nd Edition), edited by S. Stone
(World Scientific, Singapore, 1994)}
\def \tasi90{{\it Testing the Standard Model} (Proceedings of the 1990
Theoretical Advanced Study Institute in Elementary Particle Physics, Boulder,
Colorado, 3--27 June, 1990), edited by M. Cveti\v{c} and P. Langacker
(World Scientific, Singapore, 1991)}
\def \waw{XXVIII International Conference on High Energy
Physics, Warsaw, July 25--31, 1996}
\def \yaf#1#2#3#4{Yad. Fiz. {\bf#1}, #2 (#3) [Sov. J. Nucl. Phys. {\bf #1},
#4 (#3)]}
\def \zhetf#1#2#3#4#5#6{Zh. Eksp. Teor. Fiz. {\bf #1}, #2 (#3) [Sov. Phys. -
JETP {\bf #4}, #5 (#6)]}
\def \zpc#1#2#3{Zeit. Phys. C {\bf#1}, #2 (#3)}
\def \zpd#1#2#3{Zeit. Phys. D {\bf#1}, #2 (#3)}


\begin{thebibliography}{99}

\bibitem{RP} H. J. Lipkin, Y. Nir, H. R. Quinn and A. Snyder,
\prd{44}{1454}{1991}; M. Gronau, \plb{265}{389}{1991};  A. E. Snyder and H. R.
Quinn, \prd{48}{2139}{1993}. 

\bibitem{DGR} A. S. Dighe, M. Gronau, and J. L. Rosner, \prd{57}{1783}{1998}.

\bibitem{CLEOVP} CLEO \cn, M. Bishai \ite, CLEO Report No.~CLEO CONF 99-13,
presented at XIX International Symposium on Lepton and Photon Interactions at
High Energies, Stanford University, August 9--14, 1999. 

\bibitem{CLEOeta} CLEO \cn, S. J. Richichi \ite, CLEO Report No.~CLEO CONF
99-12, presented at 1999 Lepton-Photon Symposium \cite{CLEOVP}. 

\bibitem{CLEOPP} CLEO \cn, Y. Kwon \ite, CLEO Report No.~CLEO CONF 99-14,
presented at 1999 Lepton-Photon Symposium \cite{CLEOVP}.

\bibitem{CLEOasy} CLEO \cn, T. E. Coan \ite, CLEO Report No.~CLEO CONF 99-16,
presented at 1999 Lepton-Photon Symposium \cite{CLEOVP}.

\bibitem{GRL} M. Gronau, J. L. Rosner, and D. London, \prl{73}{21}{1994}.

\bibitem{FM} R. Fleischer and T. Mannel, \prd{57}{2752}{1998}.

\bibitem{GRgamma} M. Gronau and J. L. Rosner, \prd{57}{6843}{1998}.

\bibitem{NRPL} M. Neubert and J. L. Rosner, \plb{441}{403}{1998}.

\bibitem{NRPRL} M. Neubert and J. L. Rosner, \prl{81}{5076}{1998}.

\bibitem{MN} M. Neubert, \jhep{9902}{014}{1999}.  

\bibitem{GPY} M. Gronau, D. Pirjol, and T.-M. Yan, \prd{60}{034021}{1999}.

\bibitem{GP} M. Gronau and D. Pirjol, hep-ph/9902482, to be published in
Phys.~Rev.~D.

\bibitem{HHY} X.-G. He, W.-S. Hou, and K.-C. Yang, \prl{83}{1100}{1999}.

\bibitem{AKL} A. Ali, G. Kramer, and C.-D. Lu, \prd{58}{094009}{1998};
\ibj{59}{014005}{1999}.

\bibitem{HSW} W.-S. Hou, J. G. Smith, and F. W\"urthwein, National Taiwan
University report NTU-HEP-99-25, hep-ex/9910014, submitted to Phys.~Rev.~Lett. 

\bibitem{CY} H.-Y. Cheng and K.-C. Yang, hep-ph/9910291 (unpublished).

\bibitem{Falk} See, for example, the plenary talk by A. Falk at the 1999
Lepton-Photon Symposium \cite{CLEOVP}, Johns Hopkins University report, 1999,
hep-ph/9908520. 

\bibitem{Parodi} See, e.g., F. Parodi, P. Roudeau, and A. Stocchi,
hep-ex/9903063. 

\bibitem{HY} W.-S. Hou and K.-C. Yang, National Taiwan University report,
1999, hep-ph/9908202.

\bibitem{oldVP} CLEO \cn, presented by J. G. Smith at the 1997 Aspen Winter
Physics Conference on Particle Physics, Aspen, CO, January, 1997.  See also D.
M. Asner \ite, \prd{53}{1039}{1996}; K. Lingel, T. Skwarnicki, and J. G. Smith,
Ann.~Rev.~Nucl.~Part.~Sci.~{\bf 48}, 253 (1998).

\bibitem{Chau} L. L. Chau {\it et al.}, \prd{43}{2176}{1991}.

\bibitem{GHLR} M. Gronau, O. Hern\'andez, D. London, and J. L. Rosner,
\prd{50}{4529}{1994}.

\bibitem{resc} B. Blok, M. Gronau, and J. L. Rosner, \prl{78}{3999}{1997}; A.
Falk, A. L. Kagan, Y. Nir, and A. A. Petrov, \prd{57}{6843}{1998}; M. Gronau
and J. L. Rosner, \prd{58}{113005}{1998}; R. Fleischer, \plb{435}{221}{1998};
\epjc{6}{451}{1999}.

\bibitem{EWP} R. Fleischer, \zpc{62}{81}{1994}; \plb{321}{259}{1994};
\ibj{332}{419}{1994}; N. G. Deshpande and X.-G. He, \plb{336}{471}{1994};
\prl{74}{26}{1995}; N. G. Deshpande, X.-G. He, and J. Trampeti\'{c},
\plb{345}{547}{1995}; A. J. Buras and R. Fleischer, in {\it Heavy Flavours II},
edited by A. J. Buras and M. Lindner (World Scientific, Singapore, 1998),
p.~65, \art. 

\bibitem{GHLRP} M. Gronau, O. Hern\'andez, D. London, and J. L. Rosner,
\prd{52}{6374}{1995}; A. S. Dighe, M. Gronau, and J. L. Rosner,
\plb{367}{357}{1996}; \ibj{377}{325(E)}{1996}; 

\bibitem{WP} L. Wolfenstein, \prl{51}{1945}{1983}.

\bibitem{DGReta} A. S. Dighe, M. Gronau and J. L. Rosner, \prl{79}{4333}
{1997}. 

\bibitem{BPP} J. L. Rosner, \efi~99-10, hep-ph/9903543, \prd{60}{}{1999}. 

\bibitem{VPr} N. G. Deshpande and J. Trampeti\'{c}, \prd{41}{895}{1990}; see
also N. G. Deshpande in \stone, p.~587; A. DeAndrea, N. Di Bartolomeo, R.
Gatto, F. Feruglio, and G. Nardulli, \plb{320}{170}{1994}; G. Kramer, W. F.
Palmer and H. Simma, \zpc{66}{429}{1995}; D. Du and L. Guo, \zpc{75}{9}{1997}. 

\bibitem{Ciu} M. Ciuchini, R. Contino, E. Franco, G. Martinelli, and L.
Silvestrini, \np{B512}{3}{1998}. 

\bibitem{OldVP} CLEO \cn, T. Bergfeld \ite, \prl{81}{272}{1998}.

\bibitem{Oldetap} CLEO \cn, B. H. Behrens \ite, \prl{80}{3710}{1998}.

\bibitem{DGRsing} A. S. Dighe, M. Gronau and J. L. Rosner, \plb{367}{357}
{1996}; \ibj{377}{325(E)}{1996}.

\bibitem{HJLeta} H. J. Lipkin, \prl{46}{1307} {1981}; \plb{254}{247}{1991};
\plb{415}{186}{1997}; \ibj{433}{117}{1998}.

\bibitem{DD} J. L. Rosner, \efi~99-20, hep-ph/9905366, \prd{60}{}{1999}.

\bibitem{RFDH} R. Fleischer, \zpc{62}{81}{1994}; N. G. Deshpande and X.-G. He,
\plb{336}{471}{1994}. 

\bibitem{BSS} M. Bander, D. Silverman, and A. Soni, \prl{43}{242}{1979}.

\bibitem{BBNS} M. Beneke, G. Buchalla, M. Neubert, and C. Sachrajda, 
\prl{83}{1914}{1999}.

\bibitem{SW} M. Suzuki and L. Wolfenstein, \prd{60}{074019}{1999}.

\bibitem{BFC} A. J. Buras and R. Fleischer, \plb{341}{379}{1995}.

\bibitem{LSS} D. London, N. Sinha and R. Sinha, \prd{60}{074020}{1999}.

\bibitem{HJLI} H. J. Lipkin, \plb{445}{403}{1999}.

\bibitem{GRComb} M. Gronau and J. L. Rosner, \prd{59}{113002}{1999}.

\bibitem{RF} A. Buras and R. Fleischer, CERN report CERN-TH/98-319,
hep-ph/9810260 (unpublished). 

\bibitem{LKG} L. K. Gibbons, private communication.

\bibitem{Wurt} CLEO \cn, presented by Y. Gao and F. W\"urthwein at the
Meeting of the Division of Particles and Fields, American Physican Society
(DPF 99), Los Angeles, CA, 5--9 January 1999, hep-ex/9904008: $|T|^2 =
13 \pm 6.5$.

\bibitem{JLRFM} J. L. Rosner, \prd{42}{3732}{1990}.

\bibitem{Blaylock} G. Blaylock, plenary talk at the 1999 Lepton-Photon
Symposium \cite{CLEOVP}.

\bibitem{BB} G. Buchalla and A. J. Buras, \np{B548}{309}{1999}.

\bibitem{Red} G. Redlinger, presented at the Chicago Conference on Kaon
Physics, June 21--26, 1999, to be published.

\end{thebibliography}
\end{document}